\documentclass[a4paper]{article}

\usepackage{INTERSPEECH2022}

\newcommand{\matr}[1]{\mathbf{#1}} 

\title{\vspace{-9mm}PodcastMix: A dataset for separating music and speech in podcasts}
\name{Nicolás Schmidt$^1$, Jordi Pons$^2$, Marius Miron$^1$\vspace{-2mm}}
\address{$^1$ Music Technology Group, Universitat Pompeu Fabra \hspace{5mm} $^2$ Dolby Laboratories}
\email{nschmidtg@gmail.com, jordi.pons@dolby.com, marius.miron@upf.edu\vspace{-1mm}}
\begin{document}

\maketitle
\begin{abstract}
   We introduce PodcastMix, a dataset formalizing the task of separating background music and foreground speech in podcasts.
	We aim at defining a benchmark suitable for training and evaluating (deep learning) source separation models.
	To that end, we release a large and diverse training dataset based on programatically generated podcasts. 
	However, current (deep learning) models can incur into generalization issues, specially when trained on synthetic data. 
	To target potential generalization issues, we release an evaluation set based on real podcasts for which we design objective and subjective tests. 
	Out of our experiments with real podcasts, we find that current (deep learning) models may have generalization issues. Yet, these can perform competently, e.g., our best baseline separates speech with a mean opinion score of 3.84 (rating ``overall separation quality" {{from 1 to 5). The dataset and baselines are accessible online.}\vspace{-0.4mm}}
\end{abstract}

\section{Introduction}
\label{sec:intro}

A podcast is a recorded audio program comprising multiple episodes that typically contain music and speech, and is distributed over the Internet as audio files and metadata.
Spanning various themes and formats, podcasts have experienced a rise in popularity---generating 314 million dollars of profit in the US during 2017~\cite{podcasts,pedrero2019teenagers}. 
Our work investigates the feasibility of separating music and speech in podcasts, that could enable new playback possibilities and novel analytics tools that could become handy for podcast consumers, content creators, and to the podcasts industry at large.
For example, being able to separate music and speech in podcasts would allow for personalizing the volume of the background music at playback time, or for improving the accuracy of fingerprinting algorithms detecting which songs are being played~\cite{melendez2019open}. Further, separating the speech signal in podcasts may lead to the improvement of speech analytics tools---what could help with topic-based recommendation~\cite{clifton2020spotify}, or to visualize the content of a podcast. 
Hence, for a given podcast, our goal is to recover the original music and speech signals used to create the distributed mix.
We formalize the task as follows: for a given podcast mix $x=x_s+x_m$, we aim at recovering the corresponding speech $x_s$ and music $x_m$ signals.  
Thus, we frame this problem from the monaural source separation perspective and, inspired by the progress made by recent data-intensive deep learning models, we release PodcastMix---a large dataset to train, evaluate and benchmark novel algorithms that separate music and speech in podcasts.

To the best of our knowledge, we are the first to address and formalize the task of separating music and speech in podcasts. However, similar source separation efforts exist in the literature. Next, we discuss the similarities of the task we propose when compared to more consolidated ones. 
It is similar to speech denoising~\cite{rethage2018wavenet} since models are also required to be speaker independent, but we separate speech from music instead from background noise.
It is different to speech source separation~\cite{hershey2016deep,pariente2020asteroid}, because we do not individually separate speakers into different stems. We decided to jointly separate all the speakers into a single stem, since we observed that different speakers rarely \mbox{overlap in podcasts or radio shows.} This decision greatly simplifies the pipeline, and does not compromise the potential use of this technology. Yet, we acknowledge that different speakers may overlap and include this variation in our training data generation process. 
Finally, it is also related to music source separation~\cite{rafii2017musdb18,pons2021upsampling}. However, we separate spoken (not~sung) words from music. Furthermore, we consider music as a single stem containing all the instruments, including the singing voice, since we do not individually separate instruments into different stems. Importantly, some podcasts may contain vocal music as background. Hence, and due to the simplicity of obtaining vocal music, we include a significant amount of vocal music in our training data generation process.

Current state-of-the-art deep learning models require abundant training data to perform competently~\cite{rafii2017musdb18,veaux2016superseded}. Hence, one might require significant amounts of music and speech stems to successfully train a state-of-the-art model. 
We explore to programatically generate large amounts of training data {from existing music and speech datasets.} 
Programatically generated datasets are widely used in speech~\cite{hershey2016deep,paul1992design,panayotov2015librispeech,wichern2019wham}, music~\cite{miron2017monaural} and universal~\cite{kavalerov2019universal,wisdom2021s} source separation, where synthetic mixtures are derived from datasets containing individual speech or audio recordings.
However, recent source separation works report that synthetic datasets may have generalization issues~\cite{cosentino2020librimix,kadioglu2020empirical}. Specifically, a cross-dataset experiment highlighted the challenges of a waveform-based model, trained on a synthetic speech dataset, to generalize to unseen datasets~\cite{kadioglu2020empirical}. For this reason, we propose using multiple test sets for evaluation: a programatically generated one, and two out-of-distribution datasets to assess the generalization {capabilities of the models under study.}

We are solely aware of another podcast-related dataset,  
the Spotify Podcast Dataset~\cite{clifton2020spotify}, containing about 100,000 podcast episodes with speech transcripts and targeted at research on topic recognition and speech modelling~\cite{clifton2020spotify}. However, it does not contain separated music and speech stems for training source separation models.
In addition, the (audio) data is constrained to non-commercial research. To facilitate the distribution of PodcastMix, we build it out of readily accessible audio having Creative Commons licenses.

In short, we aim at facilitating research on separating music and speech in podcasts. To that end, our main contributions are:

\noindent \hspace{1mm} \textbf{--} Formalize the task of separating music and speech in podcasts, and {release a dataset to address it (sections~\ref{sec:intro} and \ref{sec:methodology}).}

\noindent \hspace{1mm} \textbf{--} Define a benchmark around the task that includes standardized training and evaluation sets {(sections \ref{sec:methodology} and \ref{sec:experiments})}---where our training set is programatically generated, two of our evaluation sets include music and speech reference stems to compute evaluation metrics, and a third evaluation set contains real podcasts (without reference stems) to run subjective tests.

\noindent \hspace{1mm} \textbf{--}  Benchmark and release two state-of-the-art models that we set as baselines for the task: Conv-TasNet~\cite{luo2019conv}, waveform-based; and U-Net~\cite{jansson2017singing}, \mbox{spectrogram-based (section \ref{sec:experiments}).}

\vspace{1mm}
\noindent Code to run our baselines and reproduce our results is on GitHub~\cite{podcastmixrepo}. The dataset is available on Zenodo~\cite{podcastmixdata}, and audio examples are also accessible online~\cite{podcastmixdemo}. Implemented using Asteroid~\cite{pariente2020asteroid} and PyTorch~\cite{paszke2019pytorch}. 

\section{PodcastMix dataset}
\label{sec:methodology}


It is composed by four sets of audio at 44.1kHz:

\noindent \hspace{1mm} \textbf{--}  \underline{PodcastMix-synth train}: large and diverse training set that is programatically generated (with a validation partition).

\noindent \hspace{1mm} \textbf{--}  \underline{PodcastMix-synth test}: a programatically generated test set with reference stems to compute evaluation metrics.

\noindent \hspace{1mm} \textbf{--}   \underline{PodcastMix-real with-reference}: a test set with real podcasts with reference stems to compute evaluation metrics.

\noindent \hspace{1mm} \textbf{--}  \underline{PodcastMix-real no-reference}: a test set with real podcasts with only podcast mixes for subjective evaluation.




\subsection{PodcastMix-synth: train and test}
\label{subsec:podcastmix_synth}

PodcastMix-synth is a programatically generated dataset imitating podcast content with high-quality Creative Commons speech and music. It comprises mono signals at 44.1kHz.

\textit{Speech signals} ---
We rely on the VCTK dataset~\cite{veaux2016superseded}, and select 44,455 short mono recordings ($\approx$41 hours) of phrases read by 110 English speakers having different accents: English, American, Irish, or Indian, among others~\cite{veaux2016superseded}.
We select recordings done using a professional microphone (DPA 4035) in a semi-anechoic chamber. Recordings were originally made at 96kHz, but we downsampled them down to 44.1kHz. 
Finally, note that other speech datasets~\cite{panayotov2015librispeech,paul1992design} could be used. Yet, we rely on VCTK data because it is a well-stablished dataset, with professional quality recordings, that allows us to operate at 44.1kHz. Further, it is sufficiently large and diverse, and is distributed using a Creative Commons license. We use 79.74\% of the recordings for training, 10.14\% for validation, and 10.12\% for testing---ensuring each partition contains different sepakers, since we aim at developing speaker-independent systems that generalize to unseen speakers.





\begin{figure}[tb]
\begin{minipage}[b]{1.0\linewidth}
  \centering
    \vspace{-1.0mm}
  \centerline{\includegraphics[width=0.82\linewidth]{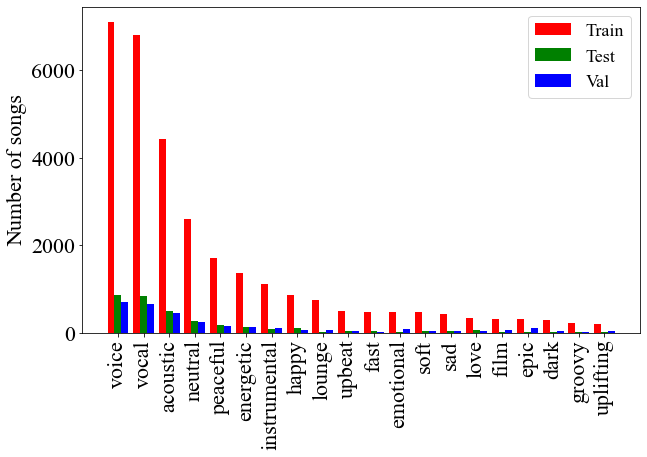}}
  \vspace{-4.0mm}
\end{minipage}
\caption{PodcastMix-synth music from Jamendo: tags.}
  \vspace{-6.0mm}
\label{fig:music-stats}
\end{figure}

\textit{Music signals} --- We rely on Jamendo as a source of Creative Commons music~\cite{bouvard2014jamendo}. 
To meet our high-quality music requirement, we focus on the $\approx$20,000 most popular songs in Jamendo, which tend to have better production quality. We use the Jamendo API to select them, which allows ranking songs by popularity (defined considering various factors including the number of downloads, plays, and likes).  We also take advantage of the Jamendo API to download stereo music with a lossless format (FLAC) at 44.1kHz. 
As a result, our music dataset is comprised of 19,370 stereo music files, from 2,763 different artists, that resemble the commercial music typically used in podcasts ($\approx$1,300 hours). 
%
%
Further, as explained in the introduction, we include a significant amount of vocal music in our training database. 
Fig.~\ref{fig:music-stats} depicts that a large portion of PodcastMix-synth contains vocal music, and that the tags distribution is similar across partitions. 
Finally, we could have relied on other Creative Commons music datasets~\cite{rafii2017musdb18,defferrard2017fma}. However,
the Jamendo tracks were the most suitable Creative Commons music we could find---e.g., the FMA dataset~\cite{defferrard2017fma} contains a fair amount of amateur experimental music that is not podcast-like.
In short, Jamendo data is of sufficient quality, large and diverse, allows us to operate at 44.1kHz, and is distributed using Creative Commons licenses.
We use 79.98\% of the recordings for training, 10.32\% for validation, and 9.70\% for testing---ensuring that each partition contains different artists.





\textit{Programatically generating podcasts --- } We generate a synthetic dataset of mono podcasts out of the above described 44,455 speech files (35,450 train / 4,507 validation / 4,498 test) and 19,370 music files (15,493 train / 1,999 validation / 1,878 test) at 44.1kHz. 
We mix them as follows:
$x=g_s x_s + g_m x_m$, where $x_s$, $x_m$ are the speech and music signals, respectively, and $g_s$, $g_m$ are their corresponding mixing gains. 
To ensure that $x_s$ sounds louder than the $x_m$ as in real podcasts ($g_s$$>$$g_m$), we scale $g_m$ by a factor $g_r={\rho(x_s)}$ $/$ ${\rho(x_m)}$, defined as the loudness ratio between speech and music. 
We compute loudness as follows: $\rho(x)=\sqrt{\sum_t{{x}^2}}$. 
%
As a result, our mixing model becomes:
$x = g_s x_s + g_r g_m x_m$, where $g_s=1$ and $g_m$ is drawn from an uniform distribution $\mathcal{U}(0.01, 1)$.  
The speech signal $x_s$ is constructed by populating an audio buffer with random speech excerpts from the same speaker. Further, in 10\% of the cases, we simulate two overlapping speakers by summing an audio excerpt from a different speaker in a random position of the buffer. 
The music signal $x_m$ is constructed by sampling non-silent fragments of the music dataset, and downmixing it to mono (via averaging the two channels) since our music signals are stereo.  
As a result, we generate convincing podcast-like data pairs $x \to (g_sx_s,\hspace{1mm}g_rg_mx_m)$ that can be used for supervised learning and evaluation. 
Hence, we now have access to a virtually infinite pool of programatically generated podcasts that are mixed on the fly (combining $\approx$41~hours of speech and $\approx$1300~hours of music) at different levels with overlapping speakers---making it suitable for training deep learning models.
Check our repository for an implementation of the proposed mixing model~\cite{podcastmixrepo} (with fixed seeds to reproduce the same dataset as ours).\footnote{We are not releasing a pre-mixed dataset to simplify its distribution, and because we do not anticipate the results to change much if the proposed mixing model is implemented slightly different (or a different random seed is used). That said, this dataset provides an interesting opportunity to investigate the influence that different data variations may have on deep learning models---and we encourage other researchers to challenge our assumption.}
Listen to some programatically generated podcasts online~\cite{podcastmixdemo}.


%






\subsection{PodcastMix-real: with-reference and no-reference}
\label{subsec:podcastmix_real}

PodcastMix-real contains real podcasts (mono at 44.1kHz) gathered to assess the generalization capabilities of the models under study. It is composed of two parts: one with reference stems to compute evaluation metrics, and another with only podcasts mixes for subjective evaluation.
These podcasts can be listened on our 
demo page~\cite{podcastmixdemo}.

\textit{PodcastMix-real with-reference} --- Provided that obtaining real high-quality Creative Commons podcasts with reference stems is difficult, we recorded our own podcast excerpts. Speech signals were recorded at 44.1kHz with a Shure SM57 microphone, including an antipop filter, plugged in into a Focusrite Scarlett 18i8 interface. The background music is from the FMA dataset~\cite{defferrard2017fma}, a Creative Commons dataset with music that is not in Jamendo. In total, we recorded 20 podcast excerpts, of 20 seconds, from 6 speakers talking 3 different languages: Portuguese, Italian, and Spanish. The recorded data pairs \mbox{$x \to (x_s,\hspace{1mm}x_m)$ can now be used to compute evaluation metrics.} 

\textit{PodcastMix-real no-reference} --- It contains 8 podcast excerpts, of 18 seconds each, that were distributed under Creative Commons licenses. These podcast mixtures, without reference stems, contain 6 different speakers talking 3 different languages: English, Chinese, and Spanish. Provided that no reference stems are available, these can be used \textit{i)} for subjective evaluations, and \textit{ii)} as a standardized set of audios to be used for sharing separations on demo websites. Hence, to facilitate comparing different models, we aim at standardizing the material that is going to be used in subjective evaluations.


\section{Experiments and benchmarking}
\label{sec:experiments}

\begin{figure}
  \hspace{7mm}\includegraphics[width=0.84\columnwidth]{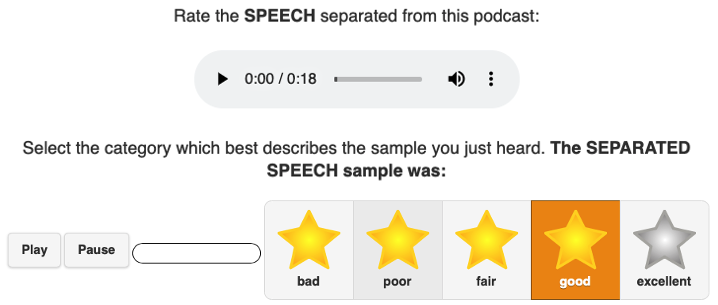}
  \vspace{-2mm}
  \caption{Subjective test: overall quality of speech separations.}
  \label{fig:test1}
    \vspace{-3mm}  
\end{figure}

We benchmark two state-of-the-art source separation models on PodcastMix: 
Conv-TasNet~\cite{luo2019conv},  a waveform-based model; and U-Net~\cite{jansson2017singing}, a spectrogram-based model.

\textit{Conv-TasNet} ---
{We adapt the Asteroid~\cite{pariente2020asteroid} implementation for 44.1kHz mono inputs}. 
Conv-TasNet~\cite{luo2019conv} consists of a learnable encoder/decoder and a separator that predicts a mask on top of this learned space.
The encoder is a CNN layer of 1024 filters (length$=$16, stride$=$4). 
The separator first uses a global layer normalization~\cite{luo2019conv} followed by a linear projection {(from 1024 to 512).}
We then use a stack of 18 blocks, repeating 3 times a combination of 6 CNN blocks with exponentially increasing dilation factors. Each CNN block is a residual layer consisting of 2~CNNs (with 512 filters interleaved with global layer normalizations and PReLUs) having skip connections~\cite{luo2019conv}. Two linear layers (with 512 filters each) adapt the dimensionality of the skip and residual connections before the next block~\cite{luo2019conv}. Finally, the outputs of the skip connections are summed and processed by an output CNN layer that predicts a mask via a ReLU.
The decoder is a tranposed CNN layer of 1024 filters (length$=$16, stride$=$4) that outputs two sources: speech and music.
Our implementation of the Conv-TasNet is online~\cite{podcastmixrepo}.

\textit{U-Net} --- We implement a U-Net~\cite{jansson2017singing} variant for 44.1kHz mono inputs.
We first compute the magnitude of the short-term Fourier transform (magnitude-STFT, with a window of 2048 samples and a hop size of 441 samples) and then use a fully-convolutional model to process it. It comprises an encoder-decoder structure with skip connections, where the encoder is composed of 6 CNN layers and the decoder is composed of 6 transposed CNN layers (each with 16, 32, 64, 128, 256, 512 filters, respectively). We use 5$\times$5 filters (stride$=$2) for both the encoder and decoder, 50\% dropout for the first 3 layers, batch normalization, LeakyReLUs, and a sigmoid activation in the last layer.
We run the magnitude-STFT input across x2 parallel U-Nets. As a result, each U-Net predicts a soft-mask: one separating the speech $\matr{M}_{s}$, and another separating the music $\matr{M}_{m}$; which we combine via $\matr{M}_{s}^{'}$ and $\matr{M}_{m}^{'}$ to filter the input mixture:
\[\matr{M}_{s}^{'} = \frac{\matr{M}_{s}}{\matr{M}_{s} + \matr{M}_{m}} \hspace{4mm} \matr{M}_{m}^{'} = \textbf{1} - \matr{M}_{s}\]
Hence, we use the phase of the mixture for reconstructing the time-domain signals. Our U-Net implementation is online~\cite{podcastmixrepo}.

\newpage

\begin{figure}
\vspace{-0mm}
  \hspace{4mm}\includegraphics[width=0.9\columnwidth]{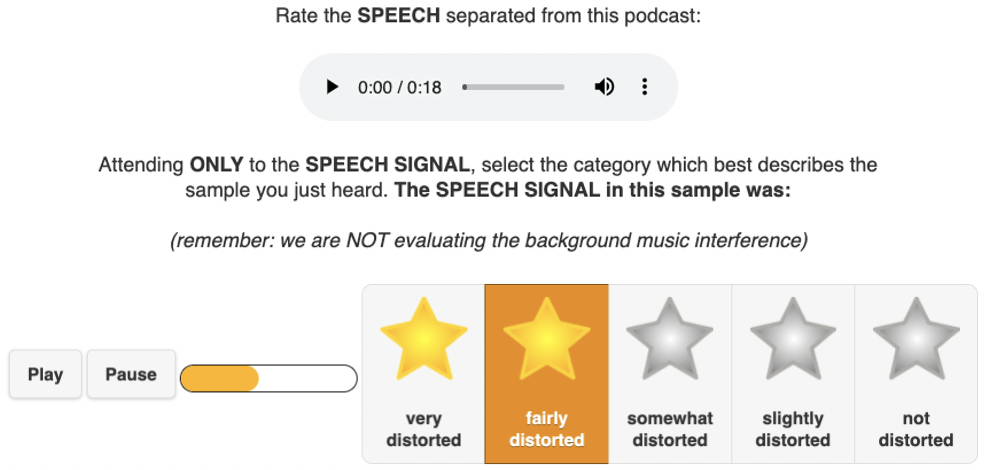}
\vspace{-2mm}
  \caption{Subjective test: distortion on speech separations.}
  \label{fig:test2}
\end{figure}

\begin{figure}
  \hspace{4mm}\includegraphics[width=0.9\columnwidth]{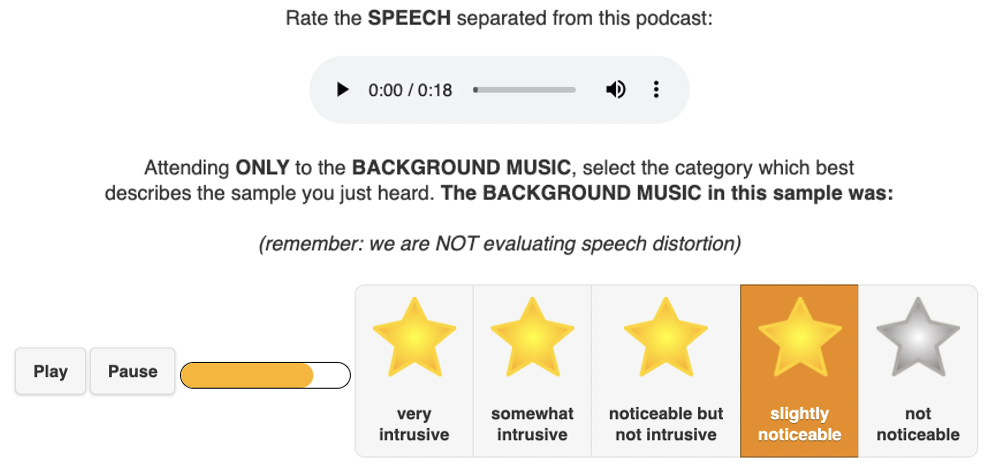}
  \vspace{-2mm}
  \caption{Subjective test: intrussiveness on speech separations.}
    \vspace{-3mm}
  \label{fig:test3}
\end{figure}

\textit{Training details} --- Conv-TasNet and U-Net were trained using the time-domain \textit{logL2} loss (related to {SI-SDR} loss \cite{heitkaemper2020demystifying}): 
\[\frac{10}{T}log10\sum_t{|\hat{x}_s(t)-x_s(t)|} + \frac{10}{T} log10\sum_t{|\hat{x}_m(t)-x_m(t)|},\]
where $\hat{x}_s(t)$ and $\hat{x}_m(t)$ are the estimated speech and music signals, and ${x}_s(t)$ and ${x}_m(t)$ are the reference stems. 
%
Models are trained for 100 epochs using the adam optimizer. Due to memory limitations, we use 2 sec inputs and a batch size of 1. The learning rate is set to 10$^{-4}$ and $\times$0.5 after the validation loss did not improve for 5 epochs.

\textit{Implementation details} --- We adjusted the Conv-TasNet and U-Net models to be comparable, of 16M parameters. Both include an adaptive input normalization, which subtracts the mean and divides by the standard deviation the podcast mixture, to perform a volume normalization that can be easily undone at the output.
Finally, note that the U-Net predicts a soft-mask---and to reconstruct the time domain signals (required for the loss above) one needs to propagate the signal (and gradients) through a differentiable inverse-STFT layer.

\begin{figure*}[t]
  \includegraphics[width=\textwidth]{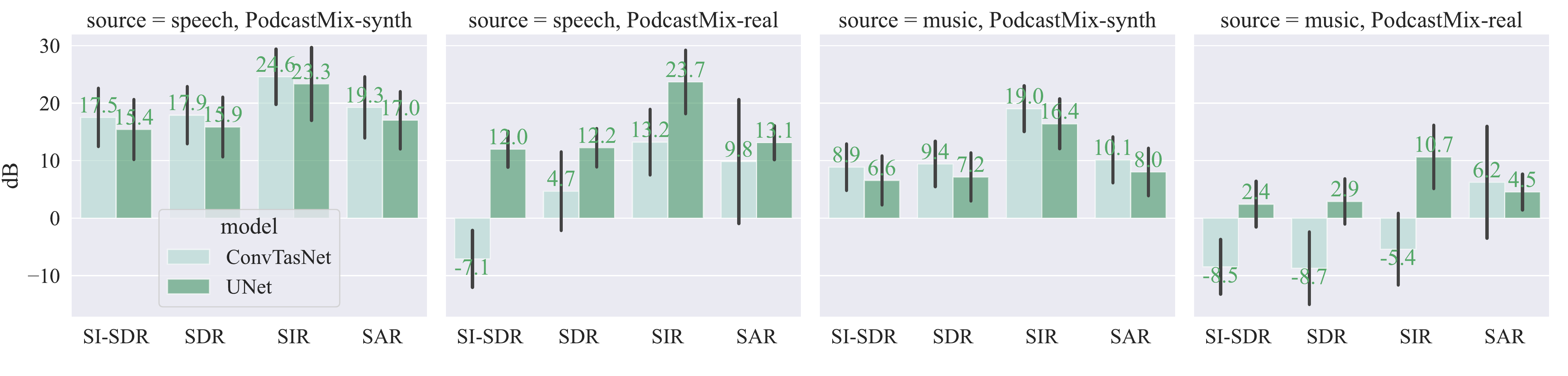}
    \vspace{-9mm}
    \caption{Evaluation metrics \cite{fevotte2005bss_eval} comparing Conv-TasNet and U-Net on `PodcastMix-synth test' and `PodcastMix-real with-reference'.}  
  \label{fig:res}
      \vspace{-2mm}
\end{figure*}

\textit{Evaluation metrics} --- We primarily rely on \textit{BSS\_eval} metrics~\cite{fevotte2005bss_eval}: source-to-distortion ratio (SDR), source-to-interference ratio (SIR), and source-to-artifact ratio (SAR). Further, as recommended by previous works~\cite{le2019sdr}, we also report scale-invariant SDR (SI-SDR). 
Since the above metrics require reference stems, we report those for PodcastMix-synth test and PodcastMix-real with-reference sets. 
Fig. \ref{fig:res} and Tables \ref{tab:speech} and \ref{tab:music} present the mean of those metrics across each of the evaluation sets we consider.
Code to compute those metrics is online~\cite{podcastmixrepo}.

\textit{Subjective evaluation} --- Provided that the above evaluation metrics can be problematic in some scenarios~\cite{cano2016evaluation}, perceptual tests are desirable when comparing source separation models. Yet, such tests can be time consuming and difficult to run. 
To facilitate and standardize subjective evaluations around PodcastMix, we use webMUSHRA~\cite{schoeffler2018webmushra}, as it allows to easily run subjective tests online; and we use 7 podcast excerpts\footnote{We use 7 (out of the 8 from PodcastMix-real no-reference) excerpts for subjective testing, since 1 is used for `training' the test participants.} from PodcastMix-real no-reference to evaluate the separated speech and music signals. 
In an attempt to further facilitate subjective testing in PodcastMix: we design a modular test comprising of two parts, and release the webMUSHRA templates we used, with audio examples~\cite{podcastmixrepo}. 
During the first part ($\approx$5 min), participants rate the overall quality (OVRL, from 1 to 5) of the separations, see Fig.~\ref{fig:test1}.
During the second part ($\approx$20 min), participants rate the separations along two axes: distortion (SIG metric, from 1 to 5, see Fig.~\ref{fig:test2}) and intrusiveness (BAK metric, from 1 to 5, see Fig.~\ref{fig:test3}).
We report mean opinion scores.
Hence, the proposed subjective evaluation follows closely the ITU-T P.835.
Note that Figs.~\ref{fig:test1},~\ref{fig:test2} and~\ref{fig:test3} only depict the subjective evaluation of speech separations, but we also run the same evaluation for music separations~(see Tables \ref{tab:speech} and \ref{tab:music}) \cite{podcastmixrepo}. Finally, we designed these experiments such that can run independently---because running a short perceptual test is preferable than not running any. 
In addition, we also split the evaluation to send the short OVRL test to a large pool of naive listeners (30 participants), and send the intricate SIG and BAK test to a selection of experienced listeners (6 participants). Mainly for 2 reasons: \textit{i)} it is easy to obtain participants for short and easy tests; and \textit{ii)} naive listeners found difficult to distinguish between distortion (SIG) and intrusiveness (BAK), for this reason we solely sent the SIG and BAK test to experienced listeners.

\textit{Results} ---
Fig.~\ref{fig:res} reports the results of our metrics-based evaluation, and Tables~\ref{tab:speech} and \ref{tab:music} also report the outcomes of our subjective evaluation. 
The audios used for the subjective test are accessible online~\cite{podcastmixdemo}.
Three main trends emerge out of the results we report:  

\noindent \hspace{1mm} \textbf{--}  The foreground speech in podcasts can be separated with \textit{good} quality by the U-Net model (OVRL of 3.84).

\noindent \hspace{1mm} \textbf{--}  Given that we target podcasts with music in the background, the music signal is more difficult to separate than speech. This is reflected across our results, which are lower for music than for speech both for Conv-TasNet and U-Net models (see SDR, OVRL and SIG scores). Further, our results also show that the speech signal is \textit{very} or \textit{somewhat intrusive} in the separated {music signals (see SIR and BAK scores)}---because in podcasts the speech signal is generally louder than music.

\noindent \hspace{1mm} \textbf{--} All the studied deep learning models have generalization issues when evaluated with real data, specially Conv-TasNet. Although better than U-Net on Podcastmix-synth test, Conv-TasNet fails to generalize with out-of-sample data. Note that Conv-TasNet obtains an OVRL$\approx$1, the worst score possible. Participants also rated the U-Net with an OVRL$\approx$3, denoting the potential of our dataset for research and benchmarking.

\noindent These are in line with previous works, where models trained with synthetic datasets had generalization issues~\cite{cosentino2020librimix,kadioglu2020empirical}, and the inductive {biases in STFT-based models helped with generalization~\cite{kavalerov2019universal,liu2021permutation}.}

\begin{table}[t]
\centering
\resizebox{\columnwidth}{!}{\begin{tabular}{c||c|c|c|c}
 & SDR ($\uparrow$) & OVRL ($\uparrow$) & SIG ($\uparrow$) & BAK ($\uparrow$) \\ \hline
U-Net & 12.2 dB & 3.84${\pm0.88}$ & 3.80${\pm0.95}$  & 4.40${\pm0.53}$   \\
Conv-TasNet & 4.7 dB & 1.18${\pm0.39}$ & 1.14${\pm0.34}$  & 3.04${\pm1.04}$
\end{tabular}}
\vspace{1mm}
\caption{Speech separation results on `PodcastMix-real' sets: SDR (with-reference) and OVRL, SIG, BAK (no-reference).\vspace{-2mm}}
\label{tab:speech}
\vspace{-4mm}
\end{table}

\begin{table}[t]
\centering
\resizebox{\columnwidth}{!}{\begin{tabular}{c||c|c|c|c}
 & SDR ($\uparrow$) & OVRL ($\uparrow$) & SIG ($\uparrow$) & BAK ($\uparrow$) \\ \hline
U-Net & 2.9 dB & 2.33${\pm0.83}$ & 3.11${\pm0.76}$  & 2.52${\pm0.62}$   \\
Conv-TasNet & -8.7 dB & 1.08${\pm0.29}$ & 4.16${\pm0.57}$ & 1.07${\pm0.25}$
\end{tabular}}
\vspace{1mm}
\caption{Music separation results on `PodcastMix-real' sets: SDR (with-reference) and OVRL, SIG, BAK (no-reference).}
\label{tab:music}
\vspace{-7mm}
\end{table}

\section{Conclusions}
\label{sec:conclusions}

\vspace{1mm}

We presented PodcastMix, a large and diverse dataset for separating music and speech in podcasts. It is divided into four parts: PodcastMix-synth (train and test), that are synthetically generated and are used for training and evaluation; and PodcastMix-real (with-reference and no-reference) that are real podcasts used for objective and subjective evaluation.
We found that it can be challenging to separate the background music from podcasts, where speech tends to be louder in the foreground. 
Although background signals are in general more difficult to separate, we want to underline that this situation is ubiquitous in podcasts---what contrasts with other source separation tasks like speech source separation (where speakers communicate at similar levels), or speech enhancement (where the background noise is not separated, but removed).
Another challenge relates to the generalization capabilities of current (deep learning) models, since we found a significant performance drop when testing our baselines with real data.
Provided that this is a recurrent topic in the literature~\cite{cosentino2020librimix,kadioglu2020empirical}, we believe that the proposed benchmark (with explicit methodologies and data to evaluate out-of-distribution behavior) can also be useful to investigate this issue.
The release of the U-Net baseline is another significant contribution. To the best of our knowledge, we are the first to release an open source model for speech and music separation in podcasts~\cite{podcastmixrepo}, and we hope to foster 
creative podcast applications, and to help improving on other tasks (e.g.,~music fingerprinting in podcasts, or podcasts recommendation).
Finally, note that PodcastMix may be easily expanded. For example, our music dataset consists of stereo signals.
Hence, a stereo version of PodcastMix is \mbox{possible---but} a (non-obvious) stereo mixing model needs to be developed. In addition, PodcastMix data may also be expanded, e.g., to separate a third \mbox{source: effects (to target music, speech and effects separation).}

\vspace{2mm}

\noindent\textbf{Acknowledgements ---}
\sloppy
{Partially funded by the Spanish Ministerio} de Ciencia, Innovación y {Universidades and the Agencia Estatal de Investigación via Musical AI {\footnotesize\&} NextCore: PID2019-111403GB-I00/AEI/10.13039/501100011033 and RTC2019-007248-7.}

\bibliographystyle{IEEEtran}
\bibliography{0_biblio}

\end{document}